\documentclass[twocolumn,preprintnumbers,prd,superscriptaddress,nofootinbib,floatfix,showpacs,showkeys]{revtex4-1}

\usepackage{amsmath}
\usepackage{amsfonts}
\usepackage{amssymb}
\usepackage{graphicx}
\usepackage{color}
\usepackage{hyperref}
\usepackage{booktabs}
\usepackage{hyperref}
\usepackage{cleveref}
\usepackage{braket}
\usepackage{mathtools}
\usepackage[rightcaption]{sidecap}
\usepackage{subfigure}

\usepackage{comment}

\definecolor{vividviolet}{rgb}{0.62, 0.0, 1.0}
\definecolor{amaranth}{rgb}{0.9, 0.17, 0.31}
\definecolor{palatinateblue}{rgb}{0.15, 0.23, 0.89}
\definecolor{brightpink}{rgb}{1.0, 0.0, 0.5}
\definecolor{cornflowerblue}{rgb}{0.39, 0.58, 0.93}
\definecolor{deepcarminepink}{rgb}{0.94, 0.19, 0.22}
\definecolor{radicalred}{rgb}{1.0, 0.21, 0.37}
\hypersetup{ linktoc=all,
    colorlinks, linkcolor={palatinateblue},
    citecolor={brightpink}, urlcolor={amaranth}
}

\newcommand{\be}{\begin{equation}}
\newcommand{\ee}{\end{equation}}
\newcommand{\bs}{\begin{split}}
\newcommand{\bea}{\begin{eqnarray}}
\newcommand{\eea}{\end{eqnarray}}

\newcommand{\bes}{\begin{subequations}}
\newcommand{\ees}{\end{subequations}}

\graphicspath{{Graphics/}}

\def\be{\begin{equation}}
\def\ee{\end{equation}}
\def\bea{\begin{eqnarray}}
\def\eea{\end{eqnarray}}

\begin{document}

\title{Unifying baryogenesis with dark matter production}

\author{Orlando Luongo}
\email{Orlando.Luongo@unicam.it}
\affiliation{Dipartimento di Matematica, Universit\`a di Pisa, Largo B. Pontecorvo, 56127 Pisa, Italy.}
\affiliation{Universit\`a di Camerino, Divisione di Fisica, Via Madonna delle carceri, 62032 Camerino, Italy.}
\affiliation{NNLOT, Al-Farabi Kazakh National University, Al-Farabi av. 71, 050040 Almaty, Kazakhstan.}

\author{Nicola Marcantognini}
\email{Nicola.Marcantognini@studenti.unicam.it}
\affiliation{Universit\`a di Camerino, Divisione di Fisica, Via Madonna delle carceri, 62032 Camerino, Italy.}

\author{Marco Muccino}
\email{Marco.Muccino@lnf.infn.it}
\affiliation{NNLOT, Al-Farabi Kazakh National University, Al-Farabi av. 71, 050040 Almaty, Kazakhstan.}
\affiliation{Istituto Nazionale di Fisica Nucleare (INFN), Laboratori Nazionali di Frascati, 00044 Frascati, Italy.}

\begin{abstract}
We here propose a mechanism that predicts, at early times, both baryon asymmetry and dark matter   origin and that recovers the spontaneous baryogenesis during the reheating. Working with $U(1)$-invariant quark $Q$ and lepton $L$ effective fields, with an interacting term that couples the evolution of Universe's environment field $\psi$, we require a spontaneous symmetry breaking and  get a pseudo Nambu-Goldstone boson $\theta$. The pseudo Nambu-Goldstone boson speeds the Universe up during inflation, playing the role of inflaton, enabling baryogenesis to occur. Thus, in a quasi-static approximation over $\psi$, we impressively find both baryon and dark matter quasi-particle production rates, unifying \emph{de facto} the two scenarios. Moreover, we outline  particle mixing and demonstrate dark matter takes over baryons.
Presupposing that $\theta$ field energy density dominates as baryogenesis stops and employing recent limits on reheating temperature, we get numerical bounds over dark matter constituent, showing that the most likely dark matter boson would be consistent with MeV-scale mass candidates. Finally, we briefly underline our predictions are suitable to explain the the low-energy electron recoil event excess between $1$ and $7$~keV found by the \texttt{XENON1T} collaboration.
\end{abstract}

\maketitle
%\tableofcontents

%%%%%%%%%%%%%%%%%%%%%%%%%%%%%%%%%%%%%%%%%%%%%%%%%%%%%%%%
%%%%%%%%%%%%%%%%%%%%%%%%%%%%%%%%%%%%%%%%%%%%%%%%%%%%%%%%

\section{Overview}

%%%%%%%%%%%%%%%%%%%%%%%%%%%%%%%%%%%%%%%%%%%%%%%%%%%%%%%%
%%%%%%%%%%%%%%%%%%%%%%%%%%%%%%%%%%%%%%%%%%%%%%%%%%%%%%%%

Current comprehension of the standard Big Bang paradigm struggled over how to fix considerable issues, above all, the \emph{cosmological constant problem} \cite{martin,wei}, a \emph{ad hoc} baryon production, named after \emph{baryogenesis} \cite{baryo}, \emph{dark matter} (DM) and \emph{dark energy} \cite{bert,revDE,revDE2}, \emph{quantum gravity} \cite{kief}, and so forth. Similarly, recent experimental tensions suggest  the Big Bang model could somehow be theoretically incomplete \cite{tensioni}.

To circumvent the problem of baryogenesis and DM production, we here conjecture a  mechanism that unifies both baryon and DM genesis under the same standards. We extend spontaneous baryogenesis considering quark $Q$ and lepton $L$ effective fields, embedded in (and coupled with the evolution) of a Universe environment field $\psi$.
All these fields possess a $U(1)$ global invariance, \textit{i.e.}, $\psi \to e^{i\alpha}\psi$, $Q \to e^{i\alpha}Q$ and $L \to L$, albeit our Lagrangian does not contain strong interactions for simplicity. We interpret the pseudo Nambu-Goldstone boson $\theta$, resulting from the baryonic symmetry breaking, as inflaton and show that two stages occur, having a first in which we claim DM to be born, whereas a second  providing a dominant barygenesis over DM. Particularly, during  reheating we recover, in a quasi-static approximation over the $\psi$ field, the abundance of baryons as expected today. We remarkably find the baryon and DM quasi-particle production rates are intertwined between them, unifying \emph{de facto} the two approaches. Further, we describe particle mixing as naive recipe to stop baryogenesis and DM production and qualitatively demonstrate why DM dominates over baryons. Assuming that the $\theta$ field energy density dominates when the baryogenesis stops and employing recent limits on the reheating temperature, we extract numerical results on the DM mass constituents, most likely congruent with MeV-scale mass candidates.

The paper is structured as follows. In Sec.~\ref{sec2} we introduce our effective model  and  in Sec.~\ref{sec3} we compute the rate of particle production for our cases and  discuss  baryogenesis, baryon asymmetry, DM production and mass mixing. The predictions of our model are also critically discussed. We highlight conclusions and perspectives of this work in Sec.~\ref{sec4}.

%%%%%%%%%%%%%%%%%%%%%%%%%%%%%%%%%%%%%%%%%%%%%%%%%%%%
%%%%%%%%%%%%%%%%%%%%%%%%%%%%%%%%%%%%%%%%%%%%%%%%%%%%
%%%%%%%%%%%%%%%%%%%%%%%%%%%%%%%%%%%%%%%%%%%%%%%%%%%%

\section{Baryogenesis in outline}\label{sec2}

%%%%%%%%%%%%%%%%%%%%%%%%%%%%%%%%%%%%%%%%%%%%%%%%%%%%%%%%
%%%%%%%%%%%%%%%%%%%%%%%%%%%%%%%%%%%%%%%%%%%%%%%%%%%%%%%%

The basic demands of our model is to get leptons formed \emph{before} baryons in order to plausibly describe baryogenesis through the  effective fields $Q$, $L$ and $\psi$ \cite{lepto}.

The evolution of the environment field $\psi$ is provided by a generalized kinetic term $\tilde X \equiv g^{\mu\nu}(\partial_\mu\bar\psi)(\partial_\nu\psi)/2$. Additionally, we build up the Dirac Lagrangian for quarks and leptons, $Q$ and $L$ with masses $m_Q$ and $m_L$, respectively,
\begin{equation}
\label{LQL}
    \mathcal{L}_{QL} = \bar Qi\gamma^\mu \partial_\mu Q - m_Q\bar QQ + \bar Li\gamma^\mu \partial_\mu L - m_L\bar LL\,.
\end{equation}
Next, we include a Lagrangian term describing the interaction between the fields $Q$, $L$ and $\psi$
\begin{equation}
\label{int}
\mathcal{L}_{\text{int}} = [i\gamma^\mu(\partial_\mu\psi)Q + h\psi\bar{L}Q + h.c.]\,,
\end{equation}
where $h.c.$ are the hermitian conjugate terms and $h$ is a coupling constant. This minimalistic choice is physically motivated by the fact that the Universe is rapidly evolving, thus, it appears natural to couple the variation of the $\psi$ field with the fermionic fields $Q$ and $L$.
In this picture, this interaction causes the reheating of the Universe.

To account for the pseudo Nambu-Goldstone boson $\theta$, playing the role of the inflaton, we resort a potential $V(\theta)$. We select two  convincing choices in agreement with the \emph{Planck} collaboration results \cite{Planck} and quadratic in $\theta$ for small oscillations around  $\theta = 0$, \textit{i.e.}, the Starobinsky \cite{staro} and the T-model \cite{Tmodel} potentials, respectively
\begin{subequations}
\begin{align}
\label{pot1}
    V_1(\theta)=&\Lambda^4\left[ 1 - \exp\left(-\sqrt{\frac{2}{3}}\frac{\psi_0\theta}{M_{\text{Pl}}}\right) \right]^2\approx \frac{2}{3}\frac{\Lambda^4\psi_0^2\theta^2}{M^2_{\text{Pl}}}\,,\\
    \label{pot2}
    V_2(\theta)= &\Lambda^4\tanh^2\left( \frac{\psi_0\theta}{ \sqrt{6\alpha}M_{\text{Pl}}}\right) \approx \frac{\Lambda^4\psi_0^2\theta^2}{6\alpha M_{\text{Pl}}^2}\,,
\end{align}
\end{subequations}
where $\Lambda$ is the amplitude and $M_{\rm Pl}$ is the Planck mass and $-2 < \log_{10}\alpha < 4$.
These choices are licit because, as we will see, the linear term $\partial_\theta V(\theta)\equiv V'(\theta)\propto\psi_0^2\theta$ enters in the equation of motion (EoM) for the $\theta$ field.
From Eqs.~\eqref{pot1}--\eqref{pot2}, we define the bare mass of the potentials as $m=\mu\Lambda^2/(\sqrt{3} M_{\text{Pl}})$ with $\mu=\{2,1/\sqrt{\alpha}\}$, respectively.

We now list below our assumptions aimed at simplifying our treatment.
\begin{itemize}
    \item[-]  The condition $h \ll 1$ ensures small enough $m_Q$ and $m_L$ so that the $\theta$ field decay produces $Q$ and $L$.
    \item[-] The $\psi$ field vacuum expectation value is $\psi = \psi_0e^{i\theta}$.
    \item[-] The $Q$ field is invariant under rotations $Q \to e^{-i\theta}Q$.
    \item[-] To avoid significant additional particle production, we assume $\partial_\mu\psi_0 \simeq 0$, which is valid as the reheating approaches its end.
\end{itemize}

Thus, implementing the above assumptions, the overall Lagrangian is given by  Eqs.~\eqref{LQL}--\eqref{int} %
\begin{align}
\nonumber
    \mathcal{L} =& \tilde X - V(\theta) + \bar Qi\gamma^\mu \partial_\mu Q - m_Q\bar QQ + \bar Li\gamma^\mu \partial_\mu L - m_L\bar LL\\
\label{Lpsi}
    &+ [h\psi_0\bar{Q}L + h.c] + \partial_\mu\theta J^\mu\,,
\end{align}
leading to the Noether baryonic current:
\begin{equation}\label{Jbar}
    J^{\mu} \equiv \bar{Q}\gamma^{\mu}Q - \psi_0\gamma^\mu(Q + \bar Q)\,.
\end{equation}
Now, we assume a spatially flat homogeneous and isotropic background, thus all the fields are functions of the time variable only.
Applying the Eulero-Lagrange equation and computing the vacuum expectation value in Heisenberg's representation, the  $\theta$ field EoM are
\begin{align}
\nonumber
   \psi_0^2(\ddot{\theta} + 3H\dot\theta) + V'(\theta) =& -i h\psi_0 \langle\bar QL - \bar LQ \rangle + ih\psi_0^2\langle L- \bar L\rangle\\
\label{theta1}
   &- i\psi_0m_Q \langle Q - \bar Q \rangle\,.
\end{align}
Solving up Eq.~\eqref{theta1} requires (a) a semiclassical approach, treating $\theta$ and $\psi_0$ as classical fields and quantizing $Q$ and $L$, and (b) a perturbative approach $\Xi(t) = \Xi_0(t) + h\Xi_1(t)$ for $h \ll 1$ \cite{Dolgov,Dolgov2}, where $\Xi_0$ generically labels the free $Q$ and $L$ fields (for $h \simeq 0$) with the condition $\dot \Xi_0 = 0$ and a vacuum expectation value $\langle \Xi_0 \rangle = 0$.

We work up to the order $h^2$ with the ansatz that the solution of the EoM of $\theta$ is a dumped oscillator with a renormalized mass $\Omega$ and amplitude assumed to vary with  time more slowly than the cosine term. The only non-zero parts on the right hand of Eq.~\eqref{theta1} are the first and third ones. Thus, the first  reads
\begin{equation}
    \langle \bar QL - \bar LQ \rangle = -\frac{ih}{4\pi}\psi_0 \Omega\dot\theta + \frac{ih}{2\pi^2}\psi_0 \Omega^2\log\left(\frac{2\omega}{\Omega}\right)\theta\,,
\end{equation}
where $\omega$ is the particle energy \cite{Dolgov}.
Instead, the third is obtained by neglecting terms proportional to $\theta\dot\theta$ or $\theta^2$ and, supposing $t$  small, $\theta t$, $\theta t^2$ or $\dot\theta t$, which would imply a still ongoing baryogenesis,
\begin{equation}
    \langle Q - \bar Q \rangle = \frac{ih^2\psi_0^3}{8\Omega^2}\left[-\dot\theta(t) + \theta(t) -\theta_i\right]\,.
\end{equation}
Finally, we substitute all our results in Eq.~\eqref{theta1}:
\begin{equation}
    \ddot\theta + (3H + \Gamma)\dot\theta + \Omega^2\theta + C = 0\,,
\end{equation}
where we defined
\begin{equation}
    \label{gamma_fric}
    \Gamma \equiv \frac{h^2 g_1^2\psi_0^2 m_Q}{8\Omega^2} + \frac{h^2}{4\pi}\Omega \quad,\quad C \equiv \frac{h^2 g_1^2\psi_0^2
   m_Q}{8\Omega^2}\theta_i\,,
\end{equation}
and qualify the renormalized mass $\Omega$ by
\begin{equation}
\label{Omega}
    m^2 \equiv \Omega^2 \left[1+ \frac{h^2 g_1^2\psi_0^2 m_Q}{8\Omega^4} + \frac{h^2}{2\pi^2}\log\left(\frac{2\omega}{\Omega}\right)\right]\,.
\end{equation}
Since $m_Q$ is negligible and $h\ll1$, it has to be $\Gamma \ll \Omega$. Assuming $H \ll \Gamma$ and applying the initial conditions $\theta(0) = \theta_i$, $\dot\theta(0) \simeq 0$, we get the damped oscillator solution
\begin{equation}\label{finaltheta}
    \theta(t) = \theta_ie^{-\Gamma t/2}\cos\left(\Omega t\right)\,.
\end{equation}

\section{Particle production}\label{sec3}

The average number density $n$ of particle-antiparticle pairs produced by our scalar field decay is formalized by \cite{Dolgov}
\begin{equation}
\label{nformula}
    n = \frac{1}{V}\sum_{s_1,s_2}\int\frac{d^3p_1}{(2\pi)^32p_1^0}\frac{d^3p_2}{(2\pi)^32p_2^0}|A|^2\,,
\end{equation}
where $A$ is the one pair production amplitude and subscripts 1 and 2 refer to the final particles produced. We need to swap it between baryons and DM for reaching baryon and DM amount of particles.

By virtue of our Lagrangian couplings, Eq.~\eqref{Lpsi} furnishes different kinds of interacting particles, comprising
\begin{itemize}
\item[{\bf 1)}] $Q\bar L$ and $\bar QL$ pairs, clearly related to the observed baryonic asymmetry \cite{Sakharov,3},
\item[{\bf 2)}] $\dot\theta Q$ and $\dot\theta\bar Q$ non-baryonic particles. Since inflatons act as source, we speculate they contribute to DM birth, leading to DM  quasi-particles\footnote{We here conjecture this mechanism can lead to a non-vanishing DM contribution. Clearly, what we here baptize \emph{DM particles} is better to say  as \emph{DM quasi-particles}. Henceforth, we only briefly name them DM particles. }. The reason why it is referred to as a source is that the corresponding Lagrangian couplings potentially generate particle excitation in the field.
\end{itemize}

\paragraph{Baryons.}

Focusing on baryons, at first order the average number density of $Q\bar L$ pairs is computed  quantizing $Q$ and $L$. Thus, perturbatively expanding over finite volumes, from Eqs.~\eqref{finaltheta} and \eqref{nformula} we get \cite{Dolgov2}
\begin{equation}
\label{n1}
    n(Q,\bar L) = \left(\frac{h\psi_0}{\pi}\right)^2\int d\omega \omega^2 \left|\int dt e^{i[2\omega t+\theta(t)]}\right|^2\,.
\end{equation}
We infer a similar expression for $n(L,\bar Q)$ with $\theta(t)$ replaced by $-\theta(t)$, with $n_b \equiv n(Q,\bar L)$  the baryon number density and $n_{\bar b} \equiv n(L,\bar Q)$ the antibaryon number density. Since $\theta$ is small, the obtainable asymmetry reads
\begin{equation}\label{barasymmetry}
    n_B = n_b - n_{\bar b} = \frac{h^2}{8\pi}\Omega\psi_0^2\theta_i^3\,.
\end{equation}

\paragraph{Dark matter.}

There is still one more part of our plan to carry out. So, working out  DM density as
\begin{equation}
    n(\dot\theta,Q) = \frac{1}{V}\sum_{s}\int\frac{d^3p}{(2\pi)^32p^0}|A_{\text{DM}}|^2\,,
\end{equation}
with amplitude
\begin{equation}
\label{ADM}
    A_{\text{DM}} = \langle Q(p, s)|\psi_0\int d^4x \dot\theta(t)\bar Q(x)e^{i\theta(t)}|0\rangle\,,
\end{equation}
so quantizing $Q$ and solving, analogously to baryons, we get
\begin{equation}
    n(\dot\theta,Q) \simeq \psi_0^2 m_Q \theta_i^2 \left(\gamma^0 + 1\right)\,,
\end{equation}
ending up with DM asymmetry:
\begin{equation}\label{dmasymmetry}
    n_{\text{DM}} = n(\dot\theta,Q) - n(\dot\theta,\bar Q) \simeq 2 \psi_0^2m_Q\theta_i^2\,.
\end{equation}

\paragraph{Mass mixing.}

Two stages occur: during the first one, the interaction term $\dot\theta\psi_0\gamma^\mu Q$ dominates in view of the large value of $\psi_0$, producing {\it de facto} the DM; during the second one, as $\dot\theta\rightarrow0$, the newly DM production becomes negligible, leaving the baryon production dominant. The mass mixing in the initial stage can be evaluated from the complete mass matrix
\begin{equation}
    M = \begin{pmatrix} m_Q & - h\psi_0 & \gamma^0\psi_0 \\ - h\psi_0 & m_L & 0 \\ \gamma^0\psi_0 & 0 & 0 \end{pmatrix}\,,
\end{equation}
so requiring real eigenvalues, letting  $\Delta m \equiv m_Q - m_L$, we get the $M$ matrix eigenstates as
\begin{equation}
\nonumber
    \Phi_1 = \dfrac{L+\epsilon_1 (Q -\dot{\theta})}{\sqrt{1+2\epsilon_1^2}}, \Phi_2 = \dfrac{Q-\epsilon_2 L+\dot{\theta}}{\sqrt{2+\epsilon_2^2}}\,,\Phi_3 = \dfrac{\dot{\theta}+\epsilon_3 Q}{\sqrt{1+\epsilon_3^2}}\,,
\end{equation}
where
\begin{subequations}
\begin{align}
    \epsilon_1 &= \dfrac{h\psi_0}{\frac{m_Q + \Delta m}{3} +\psi_0 + \sqrt{\left(\frac{\Delta m}{2}\right)^2 + (1+h^2)\psi_0^2}}\,,\\
    \epsilon_2 &= \dfrac{h\psi_0}{\frac{\Delta m - m_L}{3} + \sqrt{\left(\frac{\Delta m}{2}\right)^2 + (1+h^2)\psi_0^2}}\,,\\
    \epsilon_3 &= \dfrac{m_Q+m_L}{3\psi_0}\,.
\end{align}
\end{subequations}

Now,  baryon asymmetry is the sum of terms given by the product of a number density of produced particle/antiparticle pairs times the quark content of the pair
\begin{align}
\nonumber
n_{\rm B}^{\rm M} &= \sum_i\sum_{j\neq i}  n(\Phi_i,\bar\Phi_j)|\langle Q|\Phi_i\rangle|^2 - n(\Phi_j,\bar\Phi_i) |\langle \bar Q|\bar\Phi_i\rangle|^2\\
\label{nbmix}
&= \frac{h^2}{8\pi}\Omega\psi_0^2\theta_i^3 f(\epsilon_k)\,,
\end{align}
where each $n(\Phi_i,\bar\Phi_j)$ and $n(\Phi_j,\bar\Phi_i)$ have been computed as in Eq.~\eqref{nformula} and we arranged
\begin{equation}
\nonumber
f(\epsilon_k) \equiv \sum_{i=1}^{2}\sum_{\substack{j\neq i\\ j>i}}^{3}\xi_{ij}(\epsilon_k) \zeta_{ij}(\epsilon_k)\quad,\quad k=1,2,3\,,
\end{equation}
where, as $f(\epsilon_k)$ is smashed together, it incorporates the following definitions:
\begin{align}
\nonumber
\xi_{12}(\epsilon_k)&\equiv\left(\frac{1}{2+\epsilon_2^2}-\frac{\epsilon_1^2}{1+2\epsilon_1^2}\right)\,,\\
\nonumber
\xi_{13}(\epsilon_k)&\equiv\left(\frac{\epsilon_3^2}{1+\epsilon_3^2}-\frac{\epsilon_1^2}{1+2\epsilon_1^2}\right)\,,\\
\nonumber
\xi_{23}(\epsilon_k)&\equiv\left(\frac{\epsilon_3^2}{1+\epsilon_3^2}-\frac{1}{2+\epsilon_2^2}\right)\,,
\end{align}
\begin{align}
\nonumber
    \zeta_{12}(\epsilon_k)\equiv &  \frac{\left[(1-\epsilon_3)^2-\epsilon_1^2\epsilon_2^2(1+\epsilon_3)^2\right](1+2\epsilon_1^2)(2+\epsilon_2^2)}{\left(-1-\epsilon_1\epsilon_2+\epsilon_3-\epsilon_1\epsilon_2\epsilon_3\right)^4}\,,\\
\nonumber
    \zeta_{13}(\epsilon_k) \equiv &  \frac{\left[(1-\epsilon_3)^2(1-\epsilon_1\epsilon_2)^2-4\epsilon_1^2\epsilon_2^2\right](1+2\epsilon_1^2)(1+\epsilon_3^2)}{\left(-1-\epsilon_1\epsilon_2+\epsilon_3-\epsilon_1\epsilon_2\epsilon_3\right)^4}\,,\\
\nonumber
   \zeta_{23}(\epsilon_k)\equiv &  \frac{\left[\epsilon_1^2(1+\epsilon_3)^2(1-\epsilon_1\epsilon_2)^2-4\epsilon_1^2\right](2+\epsilon_2^2)(1+\epsilon_3^2)}{\left(-1-\epsilon_1\epsilon_2+\epsilon_3-\epsilon_1\epsilon_2\epsilon_3\right)^4}\,.
\end{align}
In asymptotic regime $\dot\theta \rightarrow0$, we get $\epsilon_1=\epsilon_2=\epsilon$ and $\epsilon_3=0$, for which spontaneous baryogenesis is recovered \cite{Dolgov,Dolgov2} and for $\Delta m = 0$ we have $\epsilon = 1$, taming the asymmetry, \textit{i.e.}, $n_B^{\rm M} = 0$.

It turns out that the here developed machinery provides DM asymmetry by
\begin{align}
\nonumber
n_{\rm DM}^{\rm M} &= \sum_in(\dot\theta,\Phi_i)|\langle \dot\theta|\Phi_i \rangle|^2 - n(\dot\theta,\bar\Phi_i)|\langle \dot\theta|\bar \Phi_i\rangle|^2\\
\label{ndmmix}
&=2 \psi_0^2m_Q\theta_i^2 \chi(\epsilon_k)\quad,\quad k=1,2,3\,,
\end{align}
where
\begin{equation}
\nonumber
\chi(\epsilon_k)\equiv  \frac{\epsilon_1^2\epsilon_2^4\left(1+2\epsilon_1^2\right)+2+\epsilon_2^2+\left(1-\epsilon_1\epsilon_2\right)^4\left(1+\epsilon_3^2\right)}{\left(-1-\epsilon_1\epsilon_2+\epsilon_3-\epsilon_1\epsilon_2\epsilon_3\right)^4}\,.
\end{equation}

Confronting Eqs.~\eqref{nbmix} and \eqref{ndmmix}, we notice $n_{\rm B}^{\rm M}\propto h^2$ whereas $n_{\rm DM}^{\rm M}$ does not depend upon $h$. Provided the interplay between predominant quantities, e.g. $\psi_0$, and negligible terms, say  $m_Q$, qualitatively, the dependence $n_{\rm B}^{\rm M}\propto h^2$, with  the prescription $h\ll1$, indicates DM might dominate over baryons. Finally mass mixing ensures that the overall process is not instantaneous and smears baryogenesis out. This is a relief, since DM contribution
could, in principle, threaten to blow up.

\subsection{Predicting dark matter candidates}

Eqs.~\eqref{nbmix} and \eqref{ndmmix} have been computed within $H\ll\Gamma$ regime, lasting for about $\Delta t\approx t\approx\Gamma^{-1}$. However,  particle production properly begins as $H\approx \Omega\gg\Gamma$, \textit{i.e.}, when  $\theta$ starts oscillating around the minimum of the potential.
In this regime, lasting $\Delta t_\star\approx t_\star\approx \Omega^{-1}$, the Universe's expansion effects turn out to be non-negligible. During reheating, the Universe behaves as matter-dominated with a scale factor $a(t)\propto t^{2/3}$.
The $\theta$ extent goes as $\propto t^{-2/3}$, whereas  we are left with baryon and  DM asymmetries between the two epochs as
\begin{subequations}
    \begin{align}
        \label{baryondensity}
        \frac{n_{\rm DM\star}^{\rm M}}{n_{\rm DM}^{\rm M}}&= \frac{\Delta t_\star}{\Delta t}\left(\frac{t}{t_\star}\right)^{4/3}\approx\left(\frac{\Omega}{\Gamma}\right)^{1/3}\,,\\
        \label{DMdensity}
        \frac{n_{\rm B\star}^{\rm M}}{n_{\rm B}^{\rm M}}&= \frac{\Delta t_\star}{\Delta t}\left(\frac{t}{t_\star}\right)^2\approx\frac{\Omega}{\Gamma}\,.
    \end{align}
\end{subequations}
Hereafter, $n_{DM}^\star$ and $n_{B}^\star$ are considered as the total asymmetries. Dividing Eq.~\eqref{DMdensity} by Eq.~\eqref{baryondensity}, we infer the unknown mass of DM constituent
\begin{equation}
\label{mdm}
    m_X\simeq \frac{h^2\Omega\theta_i}{16\pi}\left(\frac{m_p}{m_Q}\right)\left(\frac{\Omega_{DM}}{\Omega_B}\right) \frac{f(\epsilon_k)}{\chi(\epsilon_k)} \left(\frac{\Omega}{\Gamma}\right)^{2/3}\,,
\end{equation}
where we compared the produced asymmetries with the cosmic densities, \textit{i.e.}, we assumed $n_{DM}^\star\simeq \rho_{cr}\Omega_{DM}/m_X$ and $n_{B}^\star\simeq \rho_{cr}\Omega_{B}/m_p$, with $m_p$ the proton mass and $\rho_{cr}$ the current critical density, namely $\rho_{cr}\equiv 3H_0^2M_{\rm Pl}^2/(8\pi)$.

Eq.~\eqref{mdm} can be further simplified considering that:
\begin{itemize}
    \item[-] recent estimates imply  $\Omega_{DM}/\Omega_B\simeq 5$;
    \item[-] $h\ll1$ leads to $\Omega \simeq m$;
    \item[-]  Universe's energy density is dominated by $V(\theta)$, having  $H(\theta_i)=\sqrt{4\pi/3} m\psi_0\theta_i/M_{\rm Pl}$;
    \item[-]   $H(\theta_i)=\Gamma$ yelds to
    \begin{equation}
    \theta_i = \sqrt{\frac{3}{4\pi}} \frac{M_{\rm Pl}}{m\psi_0}\Gamma(h,\psi_0,m,m_Q)\,,
    \end{equation}
    where $\Gamma(h,\psi_0,m,m_Q)$ is given by Eq.~\eqref{gamma_fric}.
\end{itemize}
Plugging the above into Eq.~\eqref{mdm}, we achieve
\begin{equation}
    m_X \simeq \frac{5\sqrt{3}h^2}{32\pi^{3/2}} \frac{m_p M_{\rm Pl} m^{2/3}}{\psi_0 m_Q} \frac{f(\epsilon_k)}{\chi(\epsilon_k)} \Gamma^{1/3}\,,
\end{equation}
where $f$ and $\chi$ are functions of $h$, $m_Q$ and $m_L$, since $\epsilon_k\equiv \epsilon_k(h,m_Q,m_L)$ and $\Gamma\equiv\Gamma(h,\psi_0,m,m_Q)$.

We can now sort out the reheating temperature $T_R$,  requiring  all  relativistic species energy density,  $\rho_{rad}=(\pi^2/30)g^{\star}T_R^4$, is equal to the one estimated for $\theta$ field, namely $\rho_\theta=3H^2M_{\rm Pl}^2/(8\pi)$, at the time $t=\Gamma^{-1}$. We compute
\begin{equation}
\label{Treheating}
    T_R = \left(\frac{45}{4\pi^3g^\star}\right)^{1/4} M_{\rm Pl}^{1/2} \Gamma^{1/2}\,,
\end{equation}
where $g^\star\approx107$ is the effective numbers of relativistic degrees from all particles in thermal equilibrium with photons.
Further, from Eqs.~\eqref{nbmix} and \eqref{Treheating}, we can compute baryonic asymmetry parameter $\eta=n_{\rm B \star}^{\rm M}/s$, where entropy density is $s=(2\pi^2 g^\star/45) T_R^3$ at  reheating.

\begin{figure*}
\centering
\includegraphics[width=0.45\linewidth]{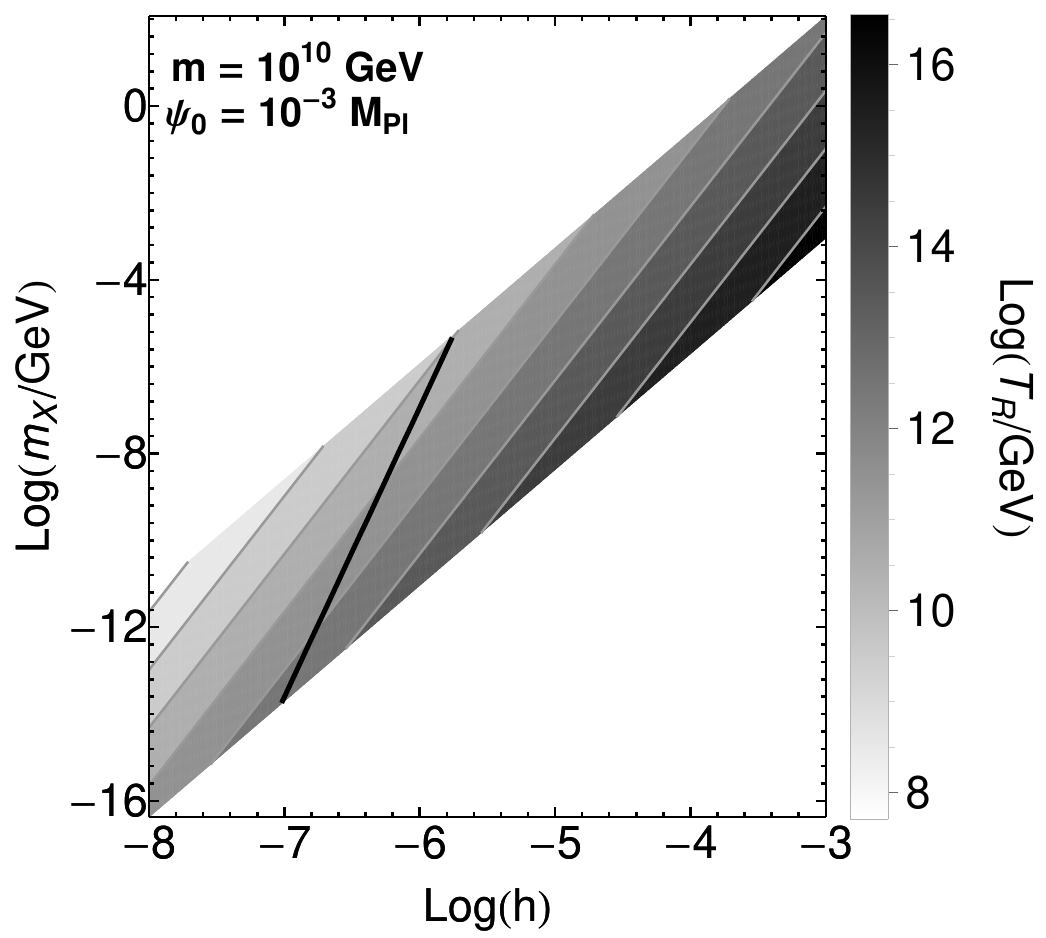}\hspace{1cm}
\includegraphics[width=0.45\linewidth]{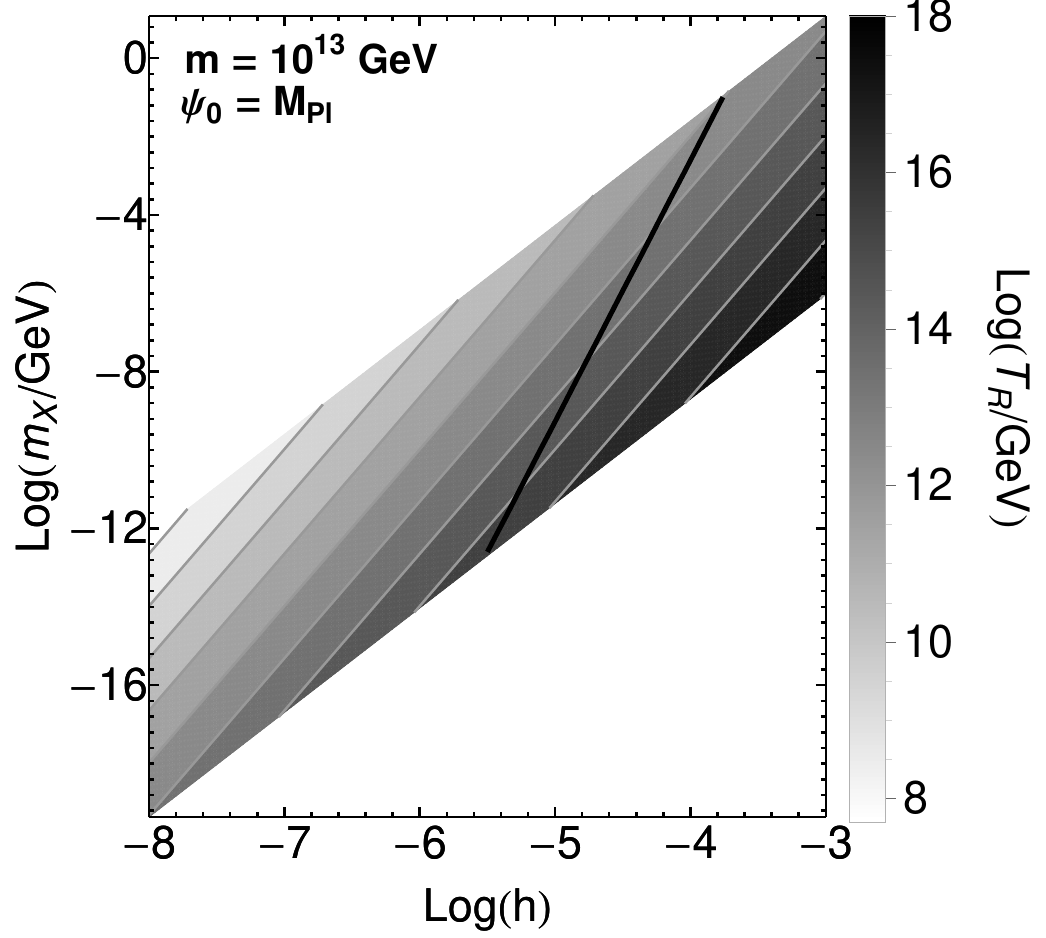}
\caption{Two dimensional contour plots $(h,m_X,T_R)$ for $\psi_0=10^{-3}M_{\rm Pl}$ and $m=10^{10}$ GeV (left panel) and for $\psi_0=M_{\rm Pl}$ and $m=10^{13}$ GeV (right panel). Solid black lines show up constraints provided by $\eta$.}
\label{fig:f1}
\end{figure*}

Thus our strategy consists in solving numerically the equations of $T_R$ and $\eta$, lumping all of information in order to get the set $(h,m_Q)$, and consequently $m_X$.
In so doing, we single out the following bounds:
\begin{itemize}
\item[-] $m\in[10^{10},10^{13}]$~GeV and $\psi_0\in[10^{-3},1]M_{\rm Pl}$ \cite{6};
\item[-] $T_R\in[10^{10},10^{15}]$~GeV \cite{6,7};
\item[-] $\eta=8.7\times10^{-11}$ \cite{Planck};
\item[-] $m_L\approx0$, having $\Delta m\approx m_Q$ and $m_t<m_Q\ll m$, where $m_t=173.2$~GeV is the top quark mass.
\end{itemize}
The numerical outcomes are displayed in Fig.~\ref{fig:f1}, where $m_X$ lower bounds are got for $m_Q= m$.
As a matter of fact,  $m_Q\ll m$ dramatically tends to shift this lower bound to a much larger value, wonderfully having thus a DM mass range within
\begin{equation}\label{results}
        10^{-5}~{\rm eV}\ll m_X \lesssim 10^7~{\rm eV}\,.
\end{equation}
On the one hand, the above DM mass rules out highly-heavy DM candidates, called variously but mostly as WIMPZillas \cite{11}.
Ultralight fields \cite{ultra}, among which axions \cite{8},
are unlikely compatible, whereas sterile neutrinos \cite{10} as fermions, look \emph{a priori} excludable  though compatible with Eq.~\eqref{results}.

On the other hand, Eq.~\eqref{results} is outstandingly compatible with tight MeV-scale DM candidates, recently proposed to explain the excess of low-energy electron recoil events between $1$ and $7$~keV measured by the \texttt{XENON1T} collaboration \cite{mev}.
Remarkably, our predictions emphasize the most reasonable DM candidate, to successfully model baryogenesis, actually lies on MeV scales.

\section{Outlooks}\label{sec4}

We here introduced a mechanism for unifying baryogenesis and DM production. We preserved spontaneous baryogenesis during  reheating, predicting the baryon asymmetry. Further, we turned our attention on how DM could form and mix, proposing DM quasi-particle owing to the couplings between pseudo Goldstone boson and quark fields.

In this respect, we set out with a $U(1)$ Lagrangian, constructed by means of effective quark $Q$ and lepton $L$ fields, with a spontaneous symmetry breaking potential, and a further interacting term that couples the evolution of Universe's environment, say $\psi_0$, with $Q$.
Immediately after the transition, the symmetry breaking potential disappeared and a pseudo Nambu-Goldstone boson, namely the inflaton field, dominated at this stage.
Within a quasi-static approximation on the environment field, we highlighted how pairs of baryons and DM particles can be produced, naively described how baryogenesis stops through the mixing process and qualitatively demonstrated why DM dominates over baryonic matter. As byproduct of our manipulations, we are therefore not tied simply to baryogenesis but the overall process yields up two sorts of massive terms, say baryons and DM. Mass mixing ensures how baryogenesis and DM production stop.

In particular, examining recent limits on $m$ and $\psi_0$ \cite{6}, $T_R$ \cite{6,7} and $\eta$ \cite{Planck}, we extracted numerical bounds on the mass range of the DM constituent, \textit{i.e.}, $10^{-5}~{\rm eV}\ll m_X \lesssim 10^7$~eV.
This result rules out highly-heavy DM candidates \cite{11} and is unlikely compatible with ultralight \cite{ultra} and axion \cite{8} fields. Remarkably, in Eq.~\eqref{results} we predicts MeV-scale mass particles as the most suitable DM candidate to successfully explain the currently observed baryonic asymmetry.
In support of, albeit independently from our prediction, such DM particles have been also proposed to explain the excess of the low-energy electron recoil events between $1$ and $7$~keV measured by the \texttt{XENON1T} collaboration \cite{mev}.

Looking ahead, in incoming works we  attempt to include quantum chromodynamics  and to unify baryogenesis with antecedent inflationary phases \cite{1}.

\acknowledgments
OL and MM acknowledge funds from the Ministry of Education and Science of the Republic of Kazakhstan,  Grant: IRN AP08052311 for financial support and are warmly thankful to A.~D.~Dolgov, C.~Freese and R.~Marotta for discussions. MM also acknowledges INFN, Frascati National Laboratories and Iniziative Specifiche MOONLIGHT2 for support. The work is dedicated to the lovely memory of B.~Luongo.

\end{document}